\documentclass[12pt]{article}

\begin{document}
\title{Lorentz surfaces with constant curvature and their physical
interpretation}
\author{Francesco Catoni, Roberto Cannata\footnote{e-mail cannata@casaccia.enea.it},
Vincenzo Catoni\footnote{e-mail vjncenzo@yahoo.it}, Paolo Zampetti\footnote{
e-mail zampetti@casaccia.enea.it}
\vspace{4mm} \\ \it ENEA; Centro Ricerche Casaccia; \vspace{1mm} \\
\it Via Anguillarese, 301;
00060  S.Maria di Galeria; Roma; Italy}
\date{July 29, 2004}
\maketitle

\def\ra{\rightarrow}
\def\Ra{\Rightarrow} \def\la{\leftarrow} \def\La{\Leftarrow}
\def\lra{\leftrightarrow} \def\Lra{\Leftrightarrow}
\newcommand{\ee}{\end{equation}}    \newcommand{\be}{\begin{equation}}
\def\ba{\begin{eqnarray}}  \def\ea{\end{eqnarray}}
\def\s{\sigma} \def\Si{\Sigma} \def\t{\tau} \def\r{\rho} \def\x{\xi}
\def\f{\phi} \def\u{\phi} \def\v{\psi} \def\c{\chi} \def\th{\theta} \def\L{\Lambda} \def\D{\Delta}
\def\z{\zeta} \def\y{\eta} \def\p{\pi} \def\o{\omega} \def\w{\omega} \def\O{\Omega}
\def\a{\alpha} \def\b{\beta} \def\g{\gamma} \def\G{\Gamma} \def\d{\delta}
\def\m{\mu} \def\l{\lambda} \def\n{\nu} \def\e{\epsilon} \def\F{\Phi}
\def\Th{\Theta} \def\sgn{\mbox{segno}} \def\2{\frac{1}{2}} \def
\Hy{\cal H}
\def\ba{\begin{eqnarray}}  \def\ea{\end{eqnarray}}
\def\2{{1\over2}}

{\bfseries SUMMARY}-
In recent years it has been recognized that the hyperbolic numbers (an extension
of complex numbers, defined as $\{z=x+h\,y; \,\,h^2=1\,\,\,x,y \in {\bf R},
\,\,h\notin {\bf R}$\}) can be associated to space-time geometry as stated by the
Lorentz transformations of special relativity. \\
In this paper we show that as the complex numbers had allowed the most
complete and conclusive mathematical formalization of the constant curvature
surfaces in the Euclidean space, in the same way the hyperbolic numbers allow
a representation of constant curvature surfaces with non-definite
line elements (Lorentz surfaces). \\
{ The results are obtained just as a consequence of
the space-time symmetry stated by the Lorentz group, but, from a
physical point of view, they give the right link
between fields and curvature as postulated by general relativity.} \\
This mathematical formalization can open new ways for application in the studies
of field theories.  

\section{Introduction}
The Riemann surfaces with constant curvature (definite line element)
form one of the essential topics of geometry, while the Lorentz surfaces
(non-definite line element) are relevant for physics \cite{Wei, KR}.
A milestone in the study of the former is the application of complex numbers
theory, that yields the best mathematical formalization \cite{Ya2, Do, Ch}.
Indeed complex analysis allows to use the most suitable
formalism, because the groups associated with complex numbers
are the same as those of the Euclidean geometry. More precisely, as
it is well known, the finite additive and unimodular multiplicative groups
of complex numbers correspond to roto-translation in the Euclidean geometry.
The infinite dimensional conformal group of functional transformations can be
fruitfully applied for the formalization of the extensions of the Euclidean 
geometry as those represented by
the non-Euclidean geometries and the differential geometry \cite[\S \ 13]{Do}. \\
The physical importance of constant curvature surfaces, represented by a non-definite
quadratic form (also called Lorentz surfaces or hyperbolic), derives from special
relativity which states that a spatial coordinate and time are linked in the
framework of a new geometry, the space-time geometry, formulated by the Lorentz
transformations of special relativity. \\
In recent years a system of numbers, previously introduced by S. Lie as a
two-dimensional example of hypercomplex numbers \cite {Lee}, has been
associated with space-time geometry \cite{Ya2}. This system of numbers, called
{\it hyperbolic} \cite{La} and defined as: $$\{z=x+h\,y; \,\,h^2=1\,\,\,x,y \in {\bf R},
 \,\,h\notin {\bf R}\}$$ could play the same role for the pseudo-Euclidean geometry as
 that played
by complex numbers in the Euclidean geometry. However, in spite of their potentialities
\cite{cz, vin}, their applications are not yet comparable with those
of complex numbers. \\ In the present work we use these numbers and the
methods of classical differential geometry for studying the surfaces
described by non-definite line elements.
We shall see that the expressions of the line elements and the equations
of the geodesics written as functions of the hyperbolic variable are the same as the
analogous expressions for constant curvature Riemann surfaces written as
functions of the complex variable. In a similar way motions on these
surfaces may be expressed by hyperbolic bilinear transformations. \\
{ If we give to the variables $x,\,y$ the physical meaning of time and
a space variable, respectively, the geodesics on constant curvature Lorentz 
surfaces represent the relativistic hyperbolic motion \cite{Ya2}.
This result has been obtained just as a consequence of the space-time symmetry
stated by Lorentz group, but it goes beyond the special relativity, because
the constant acceleration is linked to the curvature of the surfaces, in
agreement with the postulate that Einstein stated for the formalization
of the general relativity. }

The paper is organized in the following way: in sect.$\,${\bf 2} the basic
concepts on hyperbolic numbers and on geometry of the pseudo-Euclidean plane 
are briefly summarized. For a more complete introduction the reader is 
referred to \cite{Ya2, vin}. In sect.$\,${\bf 3},
by applying the results obtained in \cite{vin, ca} and by using hyperbolic
numbers, the classical results obtained for constant curvature Riemann
surfaces \cite{Do, Ch, Bi, ef} are extended to the constant curvature
Lorentz surfaces. In sect.$\,${\bf 4} a method is proposed for determining the
geodesic for two points and the geodesic distance between the same points,
as a function of the point coordinates. The method may be applied to negative
as well as positive constant curvature surfaces described by definite or
non-definite line elements. In sect.$\,${\bf 5} the physical meaning
of the obtained results is given. An appendix is added for recalling some concepts
of classical differential geometry used in this work that do not appear in
recent books.
\section{Some concepts of the pseudo-Euclidean plane geometry} \label{gpe}
\subsection{Hyperbolic numbers: basic concepts}
Here we briefly summarize some fundamental properties of hyperbolic  numbers.
This number system has been introduced by S. Lie \cite{Lee} as a two
dimensional example of the more general class of the hypercomplex number
systems. Today the hyperbolic numbers are also included in the Clifford
algebras \cite{Sok}.

Let us now introduce a hyperbolic  plane by analogy with the Gauss-Argand plane of the
complex variable. In this plane we associate the points $P\equiv(x,\,\,y)$ with
hyperbolic numbers $z=x+h\,y$.  If we represent these numbers on a Cartesian
plane, in this plane the {\it square distance} ($D$) of the point $P$ from
the origin of the coordinate axes ($O$) is defined as:
\be
D=z\tilde{z}\equiv x^2-y^2, \label{disq}
\ee
where, in analogy with complex analysis, we have indicated by $\tilde z=x-h\,y$ the
hyperbolic conjugate of $z$.\\
The definition of distance (metric element) is equivalent to introduce the
bilinear form of the {\it scalar product}. The scalar product and the properties
of hypercomplex numbers allow to state suitable axioms \cite[p. 245]{Ya2}
and to give the pseudo-Euclidean plane the structure of a vector space.

Let us consider the multiplicative inverse of $z$ that, if it exists, is given
by: $1/z\equiv \tilde z/z\tilde z.$ This implies that $z$ does not have an
inverse when $z\tilde z\equiv x^2-y^2=0$, i.e., when $y=\pm x$, or alternatively when
$z=x\pm h\,x$. In the language of algebras, these numbers are called {\it
divisors of zero}. The two straight-lines $y=\pm x$ divide the hyperbolic
plane in four sectors, and, this property is the same as that of the special
relativity representative plane and this correspondence 
 gives a physical meaning (space-time interval) to the definition of distance.
Given two points $P_j\equiv (x_j,\,y_j) ,\;P_k\equiv (x_k,\,y_k)$ we define
the {\it square distance} between them by extending
eq. (\ref{disq}):
$$D_{j,\,k}=(z_j-z_k)(\tilde z_j-\tilde z_k).$$
The hyperbolic plane has the same characteristics of the pseudo-Euclidean
flat plane \cite{cz, vin} described by the line element:
\be
ds^2=e\;(dx^2-dy^2),    \label{elenondef}
\ee
where $e=\pm 1$, in a way that $ds^2\geq 0$. \\
\subsection{Motions in the pseudo-Euclidean plane}   \label{supc10}
It is well known that motions (translations, rotations and inversions)
in the Euclidean plane are well described by complex numbers
\cite[p. 45]{Do}. In the same way motions in the pseudo-Euclidean plane
can be described by a mapping $z\ra w$ such that \cite[p. 276]{Ya2}:
\be
\mbox{ (I) }\;\;w=a\,z+b\;\;\;\;\;\;\mbox{ (II) }\;\;w=a\,\tilde z+b \;\;\;\;\;\;
(a\tilde a=1\;\Ra a=\cosh\th+h\sinh\th) \label{motion}
\ee
where $w=p+h\,q$ and $z=x+h\,y$ are hyperbolic variables and $a,\,b$ are
hyperbolic constants. Eq. (\ref{motion}, I) represents a hyperbolic
rotation about ($O$) through the hyperbolic
angle $\th$ \cite{Ya2, cz, vin} , followed by a translation by $b$.
Eq. (\ref{motion}, II) represents the same motion and a reflection in the
line $y=0 $.
It is easy to check that the mappings (I) and (II) in eq. (\ref{motion}) are
{\it square-distance-preserving}. From a geometrical point of view, because
the curves $z\tilde z=const $ represent the four arms of the equilateral
hyperbolas centered in $O$, the mappings (\ref{motion}) preserve these
hyperbolas.
Each of the above-mentioned arms is inside one of the four sectors
in which the
hyperbolic plane is divided by the straight lines (null lines)
$x\pm y=0$ \cite{Ya2,La}.
\subsection{The geodesics in the pseudo-Euclidean plane}  \label{supin01}
By means of the method summarized in appendix \ref{dcpgeo}, we can find the
geodesics in the pseudo-Euclidean plane, described by the line
element (\ref{elenondef}).
At first we have to solve the partial differential
equation with constant coefficients:
$$ \D_1\t\equiv\left(\frac{\partial\t}{\partial x}\right)^2-
\left(\frac{\partial\t}{\partial y}\right)^2=\,e\equiv\pm 1. $$ 
The elementary solution is given by $\t=Ax+By+C$ with the condition
$A^2-B^2=\pm 1$. Then we can put:
\ba
A=\cosh \th,\,\,B=\sinh\th, \qquad \mbox{ if }\qquad \D_1\t&=&+1\;; \label{cc00}\\
B=\cosh \th,\,\,A=\sinh\th, \qquad \mbox{ if }\qquad \D_1\t&=&-1\;. \label{cc11}
\ea
The equations of the geodesics are obtained equating $\partial\t/\partial\th$ to a
constant $c$, then:
\ba
\mbox{if }\qquad \D_1\t=+1& \Ra& x \sinh \th + y \cosh \th = c    \nonumber \\ 
\mbox{if }\qquad \D_1\t=-1& \Ra &x \cosh \th + y \sinh \th = c    \nonumber.   
\ea
{\it In the pseudo-Euclidean plane it is more appropriate to write the
straight line equations by means of hyperbolic trigonometric functions
instead of circular trigonometric functions as it is done in the Euclidean
plane}. \\ We note that the integration of the  equations of the
geodesics in the Euler form (see eq. \ref{eul} in App. \ref{dcpgeo}) would not give the
condition on the integration constants of eqs. (\ref{cc00}, \ref{cc11}) that
allows one to introduce the hyperbolic trigonometric functions in a natural way.

Following \cite[p. 179]{Ya2}, a segment or a straight line is said to be of
the {\it first (second) kind} if it is parallel to a line through the origin
located in the sectors containing the axis $Ox$ ($Oy$).
For a physical interpretation (special relativity), we give the $x$ variable
the physical meaning of the normalized (speed of light $c=1$) time variable
and to $y$ the physical meaning of the space variable: the lines of the first
(second) kind are called, in special relativity, {\it timelike (spacelike)}
\cite{Wei, Ya2, Na}.
As far as the hyperbolas are concerned, we may extend the definition for
the segments and straight lines and call hyperbolas of first (second) kind
those ones for which the tangent straight lines are of the first (second)
kind \cite[p. 9]{Wei}.
Following our procedure for calculating the equations of the geodesics,
we automatically obtain lines of the first kind in the case of
$\D_1\t=+1$ and lines of the second kind in the case of $\D_1\t=-1$.
\section{Constant curvature Lorentz surfaces} \label{ncc1}
\subsection{Line element} \label{31}
The sphere in three-dimensional Euclidean space and the two-sheet hyperboloid
in three-dimensional Minkowski space-time are considered as examples of
positive and negative constant curvature Riemann surfaces, respectively
\cite{Do}. \\
These results are examples of a more general theorem \cite[p. 117]{Ne},
\cite[p. 201]{ei}. Before introducing this theorem, let us point out that
in a flat space with line element
\be
ds^2=\sum^{N}_{\a=0}c_\a(dx^\a)^2 \qquad \mbox{ with }c_\a
\mbox{ constants } \label{qua}
\ee
the surfaces: $$\sum^N_{\a=0}c_\a(x^\a)^2=\pm\, R^2\,, \qquad \mbox{ with
the same }c_\a \mbox{ of eq. (\ref{qua})},$$
are called {\it fundamental hyperquadrics} \cite{Ne, ei}. For these surfaces
the following theorem holds: {\it the fundamental hyperquadrics
are $N$-dimensional spaces of constant Riemannian curvature and are the only
surfaces of constant Riemannian curvature of a $(N+1)$-dimensional flat
semi-Riemannian space.}\\
In particular, the constant curvature is positive if in the equation of the {\it fundamental
hyperquadric} there is $+R^2$ and negative if there is $-R^2$. \\
This theorem could be used for finding the line element of Lorentz
constant curvature surfaces, but here we follow a direct approach. \\
Let us consider a Lorentz surface with a line element given by:
\be
ds^2=du^2-r^2(u)\,dv^2 \label{sr}.
\ee
For surfaces with the line element given by eq. (\ref{sr})
the Gauss curvature $K$, is given by \cite{Do}:
$$K=-{{1}\over{r(u)}}{{d^2r(u)}\over{du^2}}.$$ 
In particular if we put $K=\pm R^{-2}$, with a suitable choice
of the initial conditions, we obtain: for positive
constant curvature surfaces (PCC):
\be
ds^2=du^2-R^2\sin^2(u/R)\,dv^2, \label{supc16}
\ee
 for negative constant curvature surfaces (NCC):
\be
ds^2=du^2-R^2\sinh^2(u/R)\,dv^2 \label{supc17}.
\ee
\subsection{Isometric forms of the line elements}
Following the same procedure generally used for the definite line elements
 also the non-definite line elements can be transformed
in the isometric form \cite{Do}:
\be
ds^2=f(\r,\,\f)(d\r^2-d\f^2). \label{ison1}
\ee
In particular by means of the transformation:
$$\r=-\int\frac{du}{R\sin (u/R)}\equiv  \ln \cot (u/2R);  \qquad  \f=v$$
for the line element (\ref{supc16}) and the transformation:
$$\r=-\int\frac{du}{R\sinh (u/R)}\equiv  \ln \tanh (u/2R);  \qquad  \f=v$$
for the line element (\ref{supc17}), we obtain the line elements in the
isometric form:
\be
ds^2=R^2\frac{d\r^2-d\f^2}{\cosh^2\r}\;\; \mbox{ for PCC,}  \qquad
ds^2=R^2\frac{d\r^2-d\f^2}{\sinh^2\r} \;\; \mbox{ for NCC.} \label{isop}
\ee
Moreover these isometric forms are preserved by transformations with functions of the
hyperbolic variable. In fact by introducing the hyperbolic variable,
eq. (\ref{ison1}) becomes $ds^2=g(z,\,\tilde z)dz\,d\tilde z$
and, by a hyperbolic transformation $z=F(w)$, we obtain
$ds^2=g_1 (w,\, \tilde w)|F'(w)|^2dw\,d\tilde w$ \cite{ca}. \\
Then, {\it the transformations by means of functions of the hyperbolic variable are
the conformal transformations for the non-definite differential quadratic
forms} \cite{ca}. \\
Now by means of the {\it hyperbolic exponential transformation}
\cite{vin, Fj}: 
\be
x+h\,y =R\,\exp[\r+h\,\f]\equiv R\,\exp[\r](\cosh\f+h\sinh\f), \label{a7h}
\ee
eqs. (\ref{isop}) can be rewritten in the form:
\be
ds^2=4R^4\frac{dx^2-dy^2}{(R^2+x^2-y^2)^2} \;\; \mbox{ for PCC,} \qquad 
ds^2=4R^4\frac{dx^2-dy^2}{(R^2-x^2+y^2)^2}  \mbox{ for NCC,}  \label{a8hn}
\ee
where $x,\,y$ can be considered the coordinates in a Cartesian representation.
\subsection {The equations of the geodesics}   \label{eqgeo1}
By means of the method summarized in appendix \ref{dcpgeo}, we can find the
equations of the geodesics for a surface represented by
the line element:
$$ds^2=f^2(\r)(d\r^2-d\f^2).$$  
The first step is the resolution of the partial differential equation:
$$\D_1\t\equiv\left({{\partial \t}\over{\partial \r}}\right)^2 -\left({{\partial \t}
\over{\partial \f}}\right)^2=f^2(\r).$$  
Here we follow a direct procedure, without using the general
solution given in the appendix \ref{dcpgeo}.
By the substitution
$\t=A\,\f+\t_1(\r)+C,\,\,$ where $A,\,C$ are constants,
we obtain the solution in integral form:
\be
\t=A\,\f+\int\sqrt{f^2(\r)+A^2}\;d\r +C\,.   \label{geo0h}
\ee
The equation of the geodesic is given by:
\be
{{\partial \t}\over{\partial A}}\equiv \f+\int{{A\, d\r}\over{\sqrt{f^2(\r)+
A^2}}}=B.   \label{geo1h} \\
\ee
From eqs. (\ref{geo0h}, \ref{geo1h}) we obtain a relation between $\r$ and
the line parameter $\t$:
\be
\t-A{{\partial \t}\over{\partial A}}\equiv \t-A\,B= \int {{f^2(\r)\,d\r}\over{
\sqrt{f^2(\r)+A^2}}}.   \label{geo2h}
\ee
\subsubsection{Application to constant curvature surfaces}  \label{applpccs}
Let us consider a positive constant curvature surface with the line element
given by eq. (\ref{isop}). For this case we have $f^2(\r)= R^2\cosh^{-2} \r$
and eqs. (\ref{geo0h}, \ref{geo1h}, \ref{geo2h}) become, respectively:
$$\t=A\,\f+\int\frac{\sqrt{R^2+A^2\cosh^2\r}}{\cosh \r}d\r +C $$
\ba
\sqrt{(R/A)^2+1}\sinh (B-\f) =\sinh\r \label{geos1h}\\
\t -AB=R\,\sin^{-1}\left[\frac{\tanh\r}{\sqrt{1+(A/R)^2}}\right]. \label{geos2h}
\ea
In order to obtain simplified final expressions we put in eqs. (\ref{geos1h},
\ref{geos2h}) $A=R\sinh\e,\,\,B=\s$. \\
In a similar way we obtain the equations of the geodesics for the negative constant
curvature surfaces represented by eq. (\ref{isop}). In this case we put
$A=R\sin\e,\,\,B=\s$. \\
It is well known from differential geometry \cite{Ne} that if we transform
a line element, the same transformation holds for the equations of the geodesics.
Therefore by substituting eq. (\ref{a7h}) in eq. (\ref{geos1h}) we obtain
the equations of the geodesics in a Cartesian $x,y$ plane. In fact eq.
(\ref{geos1h}) can be written $2\exp[\r](\sinh\f\cosh\s-\cosh\f\sinh\s)=
\tanh\e(\exp[2\r]-1)$. Substituting the $x,\,y$ variables from eq. (\ref{a7h})
the expressions in tab. \ref{elme} are obtained.\\
All the results are summarized in tab. \ref{elme}. In the same table the data
for constant curvature surfaces with definite line elements are also reported.
As far as the Cartesian representation is concerned we report the line
elements and the equations of the geodesics as functions of complex (definite
line elements 1-2) or hyperbolic (non-definite line elements 3-4)
variables. These expressions for definite and non-definite line elements are
the same if they are written as functions of the $z$ variable \cite{Ya2}.
\subsubsection{The {\it ``limiting curves"} }
The results reported in tab. \ref{elme} for constant curvature surfaces with
definite line element are well known and the geodesics
are represented in the $x,y$ plane by circles limited by the {\it ``limiting
circle"} \cite{Do, Bi}. The equation of the limiting circle is obtained
equating to zero
the denominator of the line element in Cartesian coordinates. For negative
constant curvature surfaces this equation is given by $x^2+y^2=R^2$, while
for positive constant curvature surfaces it is given by $x^2+y^2=-R^2$,
representing a circle with imaginary points. Then for positive constant
curvature surfaces the geodesics are complete circles. Here let us show
that the same situation applies to constant curvature surfaces with non
definite line elements. \\
For positive as well as for negative constant curvature surfaces the
geodesics are hyperbolas of the form:
\be
(y-y_0)^2-(x-x_0)^2=d^2 \label{ipgeo}
\ee
where $x_0,\,y_0,\, d$ depend on two parameters, as it can be obtained
from the equations of tab. \ref{elme}.
From the same equations we see that for $\e\ra 0$ ($A\ra 0$) the geodesics
are given by straight lines through the coordinate axes origin,
as it happens for Riemann constant curvature surfaces. \\
The expressions of $x_0,\,y_0,\, d$ as functions of $A$ and $B$ can be
obtained from the equations reported in tab. \ref{elme}. 
In particular, the half diameter $d$ of the geodesic hyperbolas, as well as
the radius of the geodesic circles on the constant curvature Riemann surfaces are:
\be
d=\frac{R^2}{A}\equiv \frac{1}{A|K|}.   \label{parge}
\ee
The limiting hyperbolas are:
$$y^2-x^2=R^2 \;\; \mbox{ for PCC;} \qquad
x^2-y^2=R^2   \;\; \mbox{ for NCC.} $$
The former limiting hyperbola does not intersect the geodesic hyperbolas. For the
latter we find the intersecting points by subtracting the equation of the
limiting hyperbola ($\F_2\equiv [x^2-y^2-R^2=0]$) multiplied by $\tan\e$
from the geodesic hyperbolas
($\F_1\equiv[(x^2-y^2+R^2)\tan\e-2R(x\sinh\s-y\cosh\s)=0]$).
In this way we have the system:
\ba
x^2-y^2-R^2=0   \nonumber   \\
R\tan\e-x\sinh\s+y\cosh\s=0  \label{inc}
\ea
Let us calculate the crossing angle between $\F_1$ and $\F_2$: the cosine
of this angle is proportional to the scalar product (in the metric of the
hyperbolic plane) of the gradients to
$\F_1$ and $\F_2$ in their crossing points:
$$ {{\partial \F_1}\over{\partial x}}
{{\partial \F_2}\over{\partial x}}-{{\partial \F_1}\over{\partial y}}
{{\partial \F_2}\over{\partial y}}=
4\,[\tan\e \,(x^2-y^2)-Rx\sinh\s+Ry\cosh\s].$$
Since in the crossing points $x^2-y^2=R^2$, due to eq. (\ref{inc}), this product
is zero. Then, as for definite line elements, $\F_1$ and $\F_2$ are
pseudo-orthogonal \cite{vin, Na, ei}. \\
We note that this property, as many others, can be considered a direct
consequence of the fact that the equilateral hyperbolas represented in the
pseudo-Euclidean plane satisfy the same theorems as the circles in the
Euclidean plane, as shown in \cite{Ya2, vin}. From an algebraic point of view this
also follows from the fact that the equations of circles and equilateral
hyperbolas are the same, if they are expressed in terms of complex and
hyperbolic variables, respectively.

\subsection{Motions}
In this subsection and in the next sect.$\,${\bf 4} we follow \cite[pag. 120]{Do}, and
use normalized variables $x/R,\,y/R\,\Ra \,x,\,y$. The line elements
of eqs. (\ref{a8hn}) become:
$$
ds^2=4R^2\frac{dx^2-dy^2}{(1+x^2-y^2)^2} \;\; \mbox{ for PCC,} \qquad 
ds^2=4R^2\frac{dx^2-dy^2}{(1-x^2+y^2)^2}  \mbox{ for NCC.}  $$
It is well known that motions (transformations that
leave unchanged the expression of the line element) on constant curvature
surfaces with
definite line element are described by means of bilinear complex
transformations \cite{Do, Bi}. Here we will show that the same result is
obtained using hyperbolic numbers for constant curvature surfaces with
non definite line elements. As it can be seen in tab. \ref{elme}, the
expressions of the line elements of the Euclidean and pseudo-Euclidean
constant curvature surfaces are the same, if they are written in
terms of complex or hyperbolic variables, respectively.
Then, also the transformations that leave unchanged the line elements are the
same, if they are written in terms of complex or hyperbolic variables.
On the other hand, one can find
directly the bilinear hyperbolic transformations that represent
motions on the constant curvature Lorentz surfaces, following the procedure
of \cite[p. 121]{Do} for the constant curvature Riemann surfaces.
In conclusion, calling $z=x+h\,y$ and $w=p+h\,q$ two hyperbolic variables and
$\a$, $\b$ two hyperbolic constants, one obtains the following
expressions for the motions:
\ba
w=\frac{\a\,z+\b}{-\tilde\b\,z+\tilde\a}\,\, ;\;\;\;
(\a\,\tilde\a+\b\,\tilde\b\neq 0) \qquad \mbox{for PCC;}    \label{movicpn} \\
w=\frac{\a\,z+\b}{\tilde\b\,z+\tilde\a}\,\,;\;\;\;
(\a\,\tilde\a-\b\,\tilde\b\neq 0) \qquad \mbox{for NCC.}    \label{movicnn}
\ea
These transformations are also reported in \cite[p. 288]{Ya2}, without
demonstration. They depend on two hyperbolic constants linked by a relation,
i.e., they actually depend on three real constants.
\section{The equation of the geodesic and geodesic distance between two points}
\label{gtzw}
Let us remind from differential geometry that two points on a surface
generally determine a geodesic and the
distance between these points can be calculated by a line integral of the
linear line element. It is known that for the constant
curvature surfaces here considered, the  equation of the geodesic as well as
the geodesic distance can be determined in an algebraic way as functions of
the point coordinates. In what follows we determine these expressions for all
the four constant curvature Lorentz and Riemann surfaces. For the Lorentz surfaces
just points for which the geodesic exists \cite[p. 150]{Ne} are considered. \\
The method we propose is based on the results already obtained, i.e., the
motions that transform geodesic lines in geodesic lines are given by
bilinear transformations. We proceed in the following way: we take the
points $P_1,\,P_2$ in the complex (hyperbolic) representative plane
$z=x+i\,y$ ($z=x+h\,y$) and look for the parameters of the bilinear
transformation
(\ref{movicpn}) or (\ref{movicnn}) that maps these points on the geodesic
straight line $q=0$ of the complex (hyperbolic) plane $w=p+i\,q$ ($w=p+h\,q$).
The inverse mapping of this straight line will give the equation of the
geodesic determined by the given points. Moreover this approach allows to
obtain the distance between two points as a function of the point coordinates. \\
As far as the distance is concerned we shall show that the proposed method
works for positive constant curvature surfaces as well as for negative ones.
\subsection{The equation of the geodesic } \label{eqgeo}
Here we consider just positive constant curvature Lorentz surfaces. Results
for the other Lorentz and Riemann constant curvature surfaces, obtained in
a similar way, are summarized in tab. \ref{5}.\\
Let us consider the two points:
$$P_1 :\, \left(z_1=x_1+h \, y_1\equiv \r_1\exp [h\th_1]\right);\,\,
P_2 :\, \left(z_2=x_2+h \, y_2\equiv \r_2\exp [h\th_2]\right)$$ and look for
the parameters ($\a,\,\b$) of the bilinear transformation that maps
$P_1,\,P_2$ in the points $P_O\equiv (0,\,0)$ and
$P_l\equiv (l,\,0)$, respectively, of the plane $w=p+h\,q$.
The $\a,\,\b$ parameters are obtained by solving the system:
$$w_1=0\Ra \a z_1=-\b;  \qquad
w_2=l\Ra\a\,z_2+\b=l(-\tilde\b\,z_2+\tilde\a).$$   
By putting $\a=\r_\a\exp[h\th_\a],\,\,\b=\r_\b\exp[h\th_\b]$ we obtain:
\ba
\r_\b\exp[h\th_\b]&=&-\r_\a\r_1\exp[h(\th_\a+\th_1)]  \label{g03} \\
(z_2-z_1)\exp[2h\th_\a]&=&l(1+\tilde z_1 z_2).         \label{g3}
\ea
Eq. (\ref{g3}) can be rewritten as:
$$l=\frac{|z_2-z_1|}{|1+\tilde z_1 z_2|}\exp\left[h\left( 2\th_\a+
\arg\left(\frac {z_2-z_1}{1+\tilde z_1 z_2}\right)\right)\right].$$   
Since $l$ is real, we obtain:
$$2\th_\a+\arg\left(\frac {z_2-z_1}{1+\tilde z_1 z_2}\right)=0,  \qquad  
l=\frac{|z_2-z_1|}{|1+\bar z_1 z_2|}.$$    
The constants of the bilinear transformations are given but for a multiplicative
constant; therefore putting $\r_\a=1$, we obtain from eq. (\ref{g03}):
\be
\th_\b=\th_\a+\th_1 ,\,\,\,\r_\b=-\r_1. \label{g4}
\ee
Now the equation of the geodesic between the points $P_1, P_2$ is derived by the
transformation of the geodesic straight line $w-\tilde w=0$ by means of eq.
(\ref{movicpn}):
\be
(x^2-y^2-1)\r_1\sinh(\th_\a+\th_\b)+x(\sinh 2\th_\a-\r_1^2\sinh 2\th_\b)+
y(\cosh 2\th_\a+\r_1^2\cosh 2\th_\b)=0.   \label{supc04}
\ee
Similarly, for the positive constant curvature surfaces with definite line
element, by putting
$\a=\r_\a\exp[i\v_\a],\,\,\b=\r_\b\exp[i\v_\b],$ eq. (\ref{g03}) becomes
$\r_\b\exp[i\v_\b]=-\r_\a\r_1\exp[i(\v_\a+\v_1)]$. Substituting, in the
right-hand side $-1=\exp[i \pi]$ we obtain:
$$ \v_\b=\pi+\v_\a+\v_1 ,\,\,\,\r_\b=\r_1.$$
The parameters of the transformations and the equations of the geodesics for
the four constant curvature surfaces are reported in tab. \ref{5}.

\subsection{Geodesic distance}        \label{supinddgg}
The transformations discussed above allow one to find the geodesic
distance $\delta(z_1,\,z_2)$ between two points $P_1$ and $P_2$ as a function
of the point coordinates.
If we take a negative constant curvature Lorentz surface,
using the line element (4) in tab. \ref{elme}, the distance between
$P_0\equiv(0,\,0)$ and $P_l\equiv(l,\,0)$ is given by
$$\delta(0,\,l)=2R\int_0^l\frac{d\,p}{1-p^2}\equiv 2R\cdot\tanh^{-1} l
\equiv R\ln\frac{1+l}{1-l}.$$   
This equation can be written as a cross ratio \cite[p. 182]{ef}, i.e., in
a form that is invariant with respect to bilinear transformations
\cite[p. 263]{Ya2}, \cite[p. 57]{Ch}. In fact we have $$\frac{1+l}{1-l}\equiv
(1,\,-1,\,0,\,l).$$
The points 1 and -1 are the intersecting points between the geodesic straight
line $q=0$ and the limiting hyperbola $x^2-y^2=1$.\\
Replacing $l$ with the expression as a function of the point coordinates,
given in tab. \ref{5}, we obtain:
\be
\delta(z_1,\,z_2)=2\,R\,\tanh^{-1}\frac{|z_1-z_2|}{|1-\bar z_1z_2|}. \label{distal}
\ee
Eq. (\ref{distal}) is also valid for negative constant curvature Riemann
surfaces.
For these surfaces the same result has been obtained in a
different way, as reported in \cite[p. 57]{Ch}. \\
These results can be extended to positive constant curvature surfaces:
{\it also for the positive constant curvature surfaces the geodesic distance
between two points is a function of a cross ratio.} \\
By means of the line element (1) of tab. \ref{elme}, we calculate the
distance between the points $P_0$ and $P_l$ for definite line element:
$$\delta(0,\,l)=2R\int_0^l\frac{d\,p}{1+p^2}\equiv 2R\arctan l\equiv\frac{R}{i}
\ln\frac{i-l}{i+l}\equiv\frac{R}{i} \ln (-i,\,i,\,0,\,l).$$   
Now $-i,\,i$ are the intersecting points between the geodesic straight
line $q=0$ and the limiting circle $x^2+y^2=-1$,
Since the properties of the cross ratio are
valid also for imaginary elements, we can again substitute for $l$ its
expression reported in tab. \ref{5}, obtaining:
\be
\delta(z_1,\,z_2)=2 R\arctan\frac{|z_1-z_2|}{|1+\bar z_1z_2|}.     \label{gep}
\ee
If we apply the same procedure for non definite line elements, we obtain the
same expression (\ref{gep}) where $z_1$ and $z_2$ are hyperbolic variables.
\section{Physical interpretation of geodesics on Riemann and Lorentz surfaces 
with positive constant curvature}
\subsection{The sphere}
The relationship existing between the curvature of bidimensional surfaces and the Laplace
equation is known \cite[p. 118]{Do}, so we can look for a physical interpretation in the
light of this equation. A geodesic circle in the $x,y$ representation
(tab. \ref{elme}, row (1)) can be considered as an equipotential curve generated by
a point source in its center. On the other hand the geodesic circles have the geometrical
meaning of stereographic projections, from the northern pole to the equatorial plane,
of the geodesic great circles on the sphere \cite[p. 96]{Do}. This projection
induces on the plane $x,y$ a Gauss metric, so that the radius of the geodesic circle
on the plane depends both on the radius of the sphere and on the position (connected
with the constant A) of the great circle on the sphere.
In fact from the data in table \ref{elme} we obtain that the
radiuses of these circles are inversely proportional to the constant $A$ times the
curvature of the starting surface: $r=1/(A\,K)$.
Then the parametric equations of these circles are given by:
\be
x=x_0 +(A\,K)^{-1}\cos [A\,K\,s], \qquad
y=y_0 +(A\,K)^{-1}\sin [A\,K\,s], \label{ce1}
\ee
where $s$ indicate the line element and $x_0,\,y_0$ the center coordinates.
\subsection{The Lorentz surfaces}
Following the positions in subsect. \ref{supin01} we give the $x$ variable
the physical meaning of a normalized (speed of light $c=1$) time variable
and $y$ the physical meaning of a space variable. The geodesics of
eq. (\ref{ipgeo}), taking into account eq. (\ref{parge}),
are given by the hyperbolas: $(x-x_0)^2-(t-t_0)^2=1/(A\,K)^2$
where $K$ is the Gauss curvature. By following the same approach used for 
writing eqs. (\ref{ce1}), we can write the equations of hyperbolas in a 
parametric form as functions of the line element:
\be
t=t_0 +(A\,K)^{-1}\sinh [A\,K\,s],\qquad
x=x_0 +(A\,K)^{-1}\cosh [A\,K\,s]. \label{hy1}
\ee
Comparing eqs (\ref{hy1}) with those of the hyperbolic motion 
\cite[p. 166]{mis}, i.e.,
$t=t_0 +(g)^{-1}\sinh g\,\t; \;\; x=x_0 +(g)^{-1}\cosh g\, \t $
($g$ is the constant acceleration), we see that the
geodesics in a plane with a constant curvature metric (eq. (\ref{hy1}))
are the same as those resulting from a motion with constant acceleration. \\
Here we note that the result of eq. (\ref{hy1}) has been obtained using the
space-time symmetry as stated by Lorentz transformations. \\
It is known that Einstein,  for formalizing the general relativity,
started from the equivalence principle and postulated that the gravitational
field would be described by the curvature tensor in a non-flat space.
Here we have obtained the relation between (gravitational) fields and the space curvature
without the need for this postulate.
\section{Conclusions}
By using the methods of differential geometry we have extended
the studies on constant curvature Riemann surfaces to constant curvature
Lorentz surfaces. In this paper we have shown that hyperbolic numbers are a
very efficient tool for studying constant curvature Lorentz surfaces, in the
same way as complex numbers are for the Euclidean space.
Then the suitable mathematics for describing space-time is the
mathematics of hyperbolic numbers \cite{cz, vin, Fj}.
Moreover the similarities between complex and hyperbolic numbers establish an
important link between the Euclidean and pseudo-Euclidean geometries.
Added to this we have shown that the obtained results are important
from a physical point of view. In fact if we give to the $x$ variable the
physical meaning of a normalized time variable, the geodesics on a
constant curvature Lorentz surfaces, represented in a Cartesian $t,\,x$ plane,
give the hyperbolic motion of special relativity. Moreover acceleration
in this motion is proportional to the surface curvature and, in this way, we
have obtained a confirmation of a postulate of the general relativity.
\appendix
\section{Differential parameter and the equations of the geodesics} \label{dcpgeo}
We recall some concepts of classical differential geometry \cite{Bi, ei}
that do not appear in recent books \cite{Ne, dC}. In particular we report
a method for obtaining the equations of the geodesics
firstly obtained by E. Beltrami
\cite[vol. I, p. 366]{Bel} who extended to the line elements the Hamilton-Jacobi
integration method, that is used for integrating the dynamics equations.
In general, it is not convenient to consider a partial differential
equation instead of the ordinary Euler equation:
\be \label{eul}
\frac{d^2 x^l}{ds^2}+\sum_{i,\,k=1}^N \G^l_{ik}\,\frac{dx^i}{ds}\,\frac{dx^k}{ds}=0,
\ee
except for particular line elements as, for instance, the examples
considered in this paper or the Schwarzschild (general relativity) metric.\\
Let us consider a N-dimensional space with a semi-Riemannian line
element given by \cite[p. 39]{ei}:
\be
ds^2=e\sum_{i,\,k=1}^N g_{ik}\,dx^i\,dx^k,    \label{elmetrgee}
\ee
where $e=\pm 1$ in a way that $ds^2\geq 0$. \\
If $\t$ is any function of the $x^i$, the function defined by:
\be
\D_1\t=\sum_{i,\,k=1}^N g^{ik}\,\frac{\partial\t}{\partial x_i}\
\frac{\partial\t}{\partial x_k}                 \label{pardif}
\ee
is called {\it Beltrami differential parameter of the first order}
\cite[p. 41]{ei}, \cite[p. 476]{Pos};
in eq. (\ref{pardif}) $g^{ik}$ are the reciprocal elements of the metric
tensor $g_{ik}$. \\
Let us now consider the non-linear partial differential equation of the first
order:
\be
\D_1 \,\t=e\,,   \label{t}
\ee
where $e$ is the same as in eq. (\ref{elmetrgee}). \\
The solution of this equation depends on an additive constant and on $N-1$
essential constants $A_i$ \cite{ei}. Now if we know the complete solution of
eq. (\ref{t}), we can obtain the equations of the geodesics by the following
theorem \cite[p. 299]{Bi}, \cite[p. 59]{ei}:
{\it when a complete solution of eq. (\ref{t}) is known, the equations of the
geodesics are given by  $\partial \t/\partial A_i=B_i$ where $A_i,\,B_i$ are
arbitrary constants, and the arc of the geodesics is given by the value of}
$\t$. \\ 
In particular if the line element has the {\it generalized Liouville} form, namely:
$$ds^2=e\,[X_1(x_1)+X_2(x_2)+...+X_N(x_N)]\sum_{i=1}^Ne_i (dx^i)^2 ,$$
where $e_i=\pm 1$, a complete integral of eq. (\ref{t}) is
$$\t=C+\sum_{i=1}^N\int\sqrt{e_i(e\,X_i+A_i)}\,dx_i,$$     
where $C$ and $A_i$ are constants, the latter being subject to the condition
$\sum_{i=1}^N A_i=0$ \cite[Vol. II, p. 426]{Bi}, \cite[p. 60]{ei}, \cite[p. 263]{dC}.
The equations of the geodesics are given, in an integral form, by:
$${{\partial \t}\over{\partial A_i}}\equiv {{1}\over{2}}\int{{e_i\,dx_i}\over{
\sqrt{e_i(e\,X_i+A_i)}}}=B_i.$$   
\newpage

\newpage

\renewcommand{\baselinestretch}{1.5}
\begin{table}[h]
\caption{Line elements and equations of the geodesics for constant curvature surfaces
with definite (1, 2) and non definite (3, 4) line elements.\protect \\ In the rows (1) and (2)
$z=x+iy$ is a complex variable obtained by means of the complex exponential transformation
of the variable $\r+i\f$; in the rows (3) and (4) $z=x+hy$ is a hyperbolic variable
obtained by means of the hyperbolic exponential transformation of the variable $\r+h\f$.
$x,\,y$ are the coordinates in the Cartesian representation; $\e,\,\s$ are
constants connected with the integration constants (A and B) of the equations of the
geodesics as it follows: $\s=B$ and
$\e=\sin^{-1}(A/R)$ in the  rows (1) and (4), $\e=\sinh^{-1}(A/R)$
in the rows (2) and (3). The constants are determined by setting two
conditions that fix the geodesic (two points or one point and the tangent in
the point). $\t$ is the linear element on the arc of geodesics. $\t_0=AB$.}
\label{elme}
\begin{center}
\begin{tabular}{|c|c|l|c|l|} \hline
\multicolumn{3}{|c|}{\emph{Isometric forms}} &
\multicolumn{2}{c|}{\emph{Cartesian isometric forms}} \\ \hline
\multicolumn{2}{|c|}{ $ds^2$}& eqs. of the geodesics &  $ds^2$& eqs. of the geodesics  \\
\hline

(1) &$ R^2{{d\r ^2+d\f^2}\over{\cosh^2 \r}}$&$ \sin (\f-\s)=\tan\e\,\sinh \r$
&$4R^4{{dx^2+dy^2}\over{(R^2+x^2+y^2)^2}}$&${{x^2+y^2}\over{R^2}}+2{{\sin\s\,x-\cos\s\,y}
\over{R\,\tan\e}}-1=0 $\\
 & &$\tanh\r=\cos\e\sin\left({{\t-\t_0}\over{R}}\right)$&$4R^4\frac{dz\,d\bar z}{(R^2+z\bar z)^2}$
 &${{z\bar{z}}\over{R^2}}+i{{z\exp[-i\s]-\bar{z}\exp[i\s]} \over{R\,\tan\e}}-1=0$ \\ \hline

(2) &$ R^2{{d\r ^2+d\f^2}\over{\sinh^2 \r}}$&$ \sin (\f-\s)=\tanh\e\,\cosh \r$
&$4R^4{{dx^2+dy^2}\over{(R^2-x^2-y^2)^2}}$&${{x^2+y^2}\over{R^2}}+2{{\sin\s\,x-\cos\s\,y}
\over{R\,\tanh\e}}+1=0 $\\
 & &$\coth\r=\cosh\e\cosh\left({{\t-\t_0}\over{R}}\right)$&$4R^4\frac{dz\,d\bar z}{(R^2-z\bar z)^2}$
 &${{z\bar{z}}\over{R^2}}+i{{z\exp[-i\s]-\bar{z}\exp[i\s]} \over{R\,\tanh\e}}+1=0$ \\ \hline

(3) &$ R^2{{d\r ^2-d\f^2}\over{\cosh^2 \r}}$&$ \sinh (\s-\f)=\tanh\e\,\sinh \r$
&$4R^4{{dx^2-dy^2}\over{(R^2+x^2-y^2)^2}}$&${{x^2-y^2}\over{R^2}}-2{{\sinh\s\,x-\cosh\s\,y}
\over{R\,\tanh\e}}-1=0 $\\
 & &$\tanh\r=\cosh\e\sin\left({{\t-\t_0}\over{R}}\right)$&$4R^4\frac{dz\,d\tilde z}{(R^2+z\tilde z)^2}$
 &${{z\tilde{z}}\over{R^2}}+h{{z\exp[-h\s]-\tilde{z}\exp[h\s]}\over{R\,\tanh\e}}-1=0$ \\ \hline

(4) &$ R^2{{d\r^2-d\f^2}\over{\sinh^2 \r}}$&$ \sinh (\s-\f)=\tan\e\,\cosh \r$
&$4R^4{{dx^2-dy^2}\over{(R^2-x^2+y^2)^2}}$&${{x^2-y^2}\over{R^2}}-2{{\sinh\s\,x-\cosh\s\,y}
\over{R\,\tan\e}}+1=0 $\\
& &$\coth\r=\cos\e\cosh\left({{\t-\t_0}\over{R}}\right)$&$4R^4\frac{dz\,d\tilde z}{(R^2-z\tilde z)^2}$
&${{z\tilde{z}}\over{R^2}}+h{{z\exp[-h\s]-\tilde{z}\exp[h\s]}\over{R\,\tan\e}}+1=0$ \\ \hline

\end{tabular}
\end{center}
\end{table}

\newpage

\begin{table}[h]
\caption{Equations of the geodesics between two points expressed as a
function of the points coordinates.\protect \\ The equations are reported for
the four constant curvature surfaces, with definite (rows (1) and (2))
and non-definite (rows (3) and (4)) line elements.
Upper part of the table:
in the rows (1) and (2): $z=x+i\,y,\,\,\bar z=x-i\,y$ are the complex variable
and its conjugate in the $x,y$ representation, $\a=\r_{\a}\exp [i\,\v_{\a}],
\,\,\b=\r_{\b}\exp [i\,\v_{\b}]$ are the constants that define the motions,
expressed in the complex polar form,
$z_1=\r_1\exp[i\,\v_1],\,\,z_2=\r_2\exp[i\,\v_2]$ are the complex coordinates
of two fixed points expressed in the complex polar form.
In the rows (3) and (4): $z=x+h\,y,\,\,\tilde z=x-h\,y$ are the hyperbolic variable
and its conjugate in the $x,y$ representation, $\a=\r_{\a}\exp [h\,\th_{\a}],
\,\,\b=\r_{\b}\exp [h\,\th_{\b}]$ are the constants that define the motions,
expressed in the hyperbolic polar form,
$z_1=\r_1\exp[h\,\th_1],\,\,z_2=\r_2\exp[h\,\th_2]$ are the hyperbolic coordinates
of two fixed points expressed in the hyperbolic polar form.
Using the parameter expressions given in the $5^{th}$, $6^{th}$ and $7^{th}$ columns
($\r_{\a}=1$), the motions given in the $2^{nd}$ column map, in the four cases,
the points of coordinate $z_1$ and $z_2$ into the points $(0,0)$ and $(l,0)$
($l$ is given in the $4^{th}$ column) of the $w$ representation.
The value of $l$ allows one to calculate the geodesic distance between two given points.
The equations of the geodesic in the $z$ $(x,y)$ representation (reported in the
lower part of the table) are obtained by transforming the axis of the abscises
of the $w$ representation.}
\label{5}
\begin{center}
\begin{tabular}{|c|c|c|c|c|c|c|} \hline
\multicolumn{3}{|c|}{$\;\;\;\;\;\;\;\;\;\;\; Motions$}&$l$&$\v_\a$,$\,\th_\a$
&$ \v_\b$,$\,\th_\b$ &$\r_\b $ \\ [0.7mm]\hline \hline
(1)&$w=\frac{\a\,z+\b}{-\bar\b\,z+\bar\a}$&$z=\frac{\bar\a\,w-\b}{\bar\b\,w
+\a}$&$\;\;\;\;\frac{|z_2-z_1|}{|1+\bar{z}_1
\,z_2|}\;\;\;\;$&${{1}\over{2}}\arg\left({{1+\bar{z}_1\,z_2}\over{z_2-z_1}}
\right)$&$\pi+\v_\a+\v_1 $&$\r_1 $ \\ [1.2mm]\hline
(2)&$w=\frac{\a\,z+\b}{\bar\b\,z+\bar\a}$&$z=\frac{-\bar\a\,w+\b}{\bar\b\,w-\a}
$&$\frac{|z_2-z_1|}{|1-\bar{z}_1\,z_2|}$&
${{1}\over{2}}\arg\left({{1-\bar{z}_1\,z_2}\over{z_2-z_1}}\right)$&
$\pi+\v_\a+\v_1 $&$\r_1$ \\ [1.2mm] \hline
(3)&$w=\frac{\a\,z+\b}{-\tilde\b\, z+\tilde\a}$&$z=\frac{\tilde\a\,w-\b}
{\tilde\b\,w+\a}$&$\frac{|z_2-z_1|}{|1+\tilde{z}_1\,z_2|}$&${{1}\over{2}}\arg
\left({{1+\tilde{z}_1\,z_2}\over{z_2-z_1}}\right)$&$\th_\a+\th_1 $&$-\r_1$
 \\ [1.2mm] \hline
(4)&$w=\frac{\a\,z+\b}{\tilde\b\,z+\tilde\a}$&$z=\frac{-\tilde\a\,w+\b}{
\tilde\b\,w-\a}$& $\frac{|z_2-z_1|}{|1-\tilde{z}_1\,z_2|}$&${{1}\over{2}}\arg
\left({{1-\tilde{z}_1\,z_2}\over{z_2-z_1}}\right)$&$\th_\a+\th_1 $&$
-\r_1$ \\ [1.2mm] \hline\hline
\multicolumn{7}{|c|}{The equations of the geodesics} \\ [1mm] \hline
(1)& \multicolumn{6}{|c|}{
$(x^2+y^2-1)\r_1\sin(\v_\a+\v_\b)-x(\sin 2\v_\a-\r_1^2
\sin 2\v_\b)-y(\cos 2\v_\a+\r_1^2\cos 2\v_\b)=0 $} \\ [0.5mm]
(2)& \multicolumn{6}{|c|}{ $(x^2+y^2+1)\r_1\sin(\v_\a+\v_\b)+x(\sin 2\v_\a+\r_1^2
\sin 2\v_\b)+y(\cos 2\v_\a-\r_1^2\cos 2\v_\b)=0 $} \\ [0.5mm]
(3)& \multicolumn{6}{|c|}{$(x^2-y^2-1)\r_1\sinh(\th_\a+\th_\b)+x(
\sinh 2\th_\a-\r_1^2
\sinh 2\th_\b)+y(\cosh 2\th_\a+\r_1^2\cosh 2\th_\b)=0 $} \\ [0.5mm]
(4)& \multicolumn{6}{|c|}{
$(x^2-y^2+1)\r_1\sinh(\th_\a+\th_\b)-x(\sinh 2\th_\a+\r_1^2
\sinh 2\th_\b)-y(\cosh 2\th_\a-\r^2_1\cosh 2\th_\b)=0 $} \\ [0.5mm]  \hline
\end{tabular}
\end{center}
\end{table}

\end{document}